# Landau theory applied to phase transitions in calcium orthotungstate and isostructural compounds

Daniel Errandonea[†]

Departamento de Física Aplicada-ICMUV, Universitat de València,

Edificio de Investigación, c/Dr. Moliner 50, 46100 Burjassot (Valencia), Spain.

The pressure-driven tetragonal-to-monoclinic phase transition in $CaWO_4$ and related scheelite-structured orthotungstates is analysed in terms of spontaneous strains. Based upon our previous high-pressure x-ray diffraction results and the Landau theory, it is suggested that the scheelite-to-fergusonite transition is of second order in nature.

---

[†] electronic mail: daniel.errandonea@uv.es,  Tel.: (34) 96 354 4475, FAX: (34) 96 354 3146

**I. Introduction**

Scheelite structured orthotungstates are technologically important materials that can be used as scintillators [1], laser-host materials [2], or cryogenic detectors for dark matter [3]. Several authors have investigated the effects of pressure on the properties of the AWO$_4$ compounds: CaWO$_4$, SrWO$_4$, BaWO$_4$, PbWO$_4$, and EuWO$_4$ [4 – 24]. All these compounds crystallize at atmospheric pressure in a tetragonal structure (space group: I4$_1$/$a$, No. 88, Z = 4) which is isostructural with the mineral scheelite (CaWO$_4$) [25]; see Fig. 1. In this structure, the A and W sites have an S$_4$ point symmetry, being the W atoms coordinated by four O atoms forming nearly regular tetrahedra and the A cations by eight O atoms forming bisdisphenoids [26]. It has been known since the eighties that under compression these orthotungstates undergo a phase transition from the tetragonal scheelite structure to a monoclinic structure [6 – 8]. However, only recently it has been possible to accurately characterize the crystalline structure of the high-pressure phase, being assigned to a monoclinic structure (space group: I2/$a$, No. 15, Z = 4) [13, 15 - 18, 21] which is isostructural with the mineral fergusonite (YNbO$_4$) [27]; see Fig. 1. The pressure-induced phase transition in AWO$_4$ compounds, from scheelite to fergusonite or its reverse, occurs in the range of 7 GPa to 11 GPa [16, 17, 21]. An analogous transition, but induced by temperature, is known to take place in LaNbO$_4$ [28]. This phase transformation has been widely studied, being characterized as a second-order transition from a high-temperature paraelastic state (scheelite-type phase) to a low-temperature ferroelastic state (fergusonite-type phase) [29]. In the majority of materials of the ABO$_4$-type (particularly in LaNbO$_4$), the structural changes that occur upon compression are similar to those that occur on cooling from high temperature [30]; namely there is an inverse relationship between pressure and temperature. One of the few significant exceptions to this systematic behaviour is BiVO$_4$ [31]. Based upon the above described facts, one can speculate that the pressure-induced phase transition in the scheelite-structured orthotungstates is also a ferroelastic transformation. Surprisingly, the mechanism of this pressure-driven phase transformation still remains uncertain. The aim of this work is to improve the present understanding of the scheelite-to-fergusonite transition in AWO$_4$ compounds by analyzing the spontaneous strains of the monoclinic phase, extracted from our previously reported x-ray diffraction experiments, on the light of the Landau theory [32].

**II. Experimental background**

The monoclinic and spontaneous strains here reported were calculated from high-pressure x-ray powder diffraction data measured at the 16-IDB beamline of the HPCAT facility at the Advanced Photon Source or at the X-17C beamline of the National Synchrotron Light Source using a diamond-anvil cell. Silicone oil was used as pressure-transmitting medium in all the experiments with the exception of those where EuWO$_4$ was studied. In this latter case a 4:1 methanol-ethanol mixture was employed as pressure-transmitting medium. A detailed description of the experiments was given in Refs. [16, 17, 21]. There, we reported the occurrence of the scheelite-to-fergusonite phase transition in CaWO$_4$, SrWO$_4$, BaWO$_4$, PbWO$_4$, and EuWO$_4$ as well as additional pressure-induced structural changes. However, in our previous works, we did not analyse into detail the mechanism driving the pressure-induced tetragonal-to-monoclinic phase transformation in this class of compounds. In the present paper we report a detailed analysis of this issue based upon the pressure evolution of the spontaneous strains in the fergusonite phase.

**III. Structural model for the scheelite-to-fergusonite phase transition**

Fergusonite is a distorted and compressed version of scheelite obtained by a small distortion of the cation matrix and significant displacements of the anions. As a matter of fact, both structure-types contain isolated $WO_4$ tetrahedra interlinked by A ions which have primarily eightfold oxygen coordination. Fig. 1 illustrates how subtle the actual change is at the transition pressure ($P_T$). Indeed, at pressures slightly higher than the transition pressure the **b** angle of the high-pressure monoclinic unit cell of the $AWO_4$ family is only a little above 90º. Moreover the *a* and *c* lattice parameters are not widely different in value [16, 17, 21]. In spite of the similarities between both structures, the slight differences between them can be still detected in Fig. 1 by comparing the relative position between the W atoms located on the right upper corner of both structures and between the A atoms located at their centers. Basically, the scheelite-to-fergusonite transition in $AWO_4$ compounds is caused by small displacements of the A atoms from their high-symmetry positions. This structural instability would bring about changes in the O positions and the consequently polyhedra distortion (see Fig. 1). Because of these atomic displacements, immediately after the transition the volume of $WO_4$ tetrahedra is enlarged by less than 10% and the volume of the $AO_8$ bisdisphenoids is reduced by a similar amount [16, 21]. In addition, the monoclinic distortion of fergusonite continuously increases upon compression enhancing these atomic displacements [16, 17, 21]. On the other hand, the scheelite-to-fergusonite transition in $ABO_4$ compounds is of martensitic nature [33, 34], being the initial structure partially conserved while certain sheets of it are slightly shifted [14]. In the case of the orthotungstates here studied, it involves a shift in the zigzag chains of W cations either along [100] or [010] directions. As a consequence of the above described atomic displacements, the discussed phase transition involves a lowering of the point-group symmetry from 4/m to 2/m, which results in two possible orientation states ($S_1$ and $S_2$) in the monoclinic phase. These two states are crystallographically equivalent.

**IV. Spontaneous strains**

In a ferroelastic transformation the $S_1$ and $S_2$ states can be seen as a small distortion caused by slight displacements of the atoms of the parent phase. The spontaneous strain characterizes the distortion of each orientation state relative to the prototype structure (i.e. the scheelite-type structure). The second rank strain tensor for monoclinic symmetry for a single orientation state ($S_1$) is given by:

$$\mathbf{e}_{ij}(S_1) = \begin{pmatrix} e_{11} & e_{12} & 0 \\ e_{12} & e_{22} & 0 \\ 0 & 0 & e_{33} \end{pmatrix}; \qquad (1)$$

and $\mathbf{e}_{ij}(S_2)$ is related to $\mathbf{e}_{ij}(S_1)$ by $\mathbf{e}_{ij}(S_2) = \mathbf{R}\mathbf{e}_{ij}(S_1)\mathbf{R}^\mathbf{T}$, where $\mathbf{R}$ and $\mathbf{R}^\mathbf{T}$ are the 90° rotation matrix around the *b*-axis of the monoclinic unit cell and its transpose. It is important to note here that the crystallographic settings used to describe the scheelite and fergusonite phases are related in such a way that the *c*-axis of the tetragonal unit cell corresponds to the *b*-axis of the monoclinic unit cell (see Fig. 1). Following Schlenker [35] the strain components can be calculated as follows:

$$e_{11} = \frac{c_M \sin b_M}{a_T} - 1, \qquad (2)$$

$$e_{22} = \frac{a_M}{a_T} - 1, \qquad (3)$$

$$e_{33} = \frac{b_M}{c_T} - 1, \qquad (4)$$

$$\text{and } e_{12} = \frac{1}{2} \frac{c_M \cos b_M}{a_T}; \qquad (5)$$

where $a_M$, $b_M$, $c_M$, and $\boldsymbol{b}_M$ are the monoclinic lattice parameters and $a_T$ and $c_T$ are the tetragonal lattice parameters. According to Aizu [36] in the present case the spontaneous strain tensor can be expressed as:

$$\mathbf{e}^s_{ij}(S_1) = \begin{pmatrix} -u & v & 0 \\ v & u & 0 \\ 0 & 0 & 0 \end{pmatrix} \text{ and } \mathbf{e}^s_{ij}(S_2) = \begin{pmatrix} u & -v & 0 \\ -v & -u & 0 \\ 0 & 0 & 0 \end{pmatrix}; \quad (6)$$

where $u = \frac{1}{2}(e_{22} - e_{11})$ is the longitudinal spontaneous strain and $v = e_{12}$ is the shear spontaneous strain. The scalar spontaneous strain $\mathbf{e}_s$ is defined as:

$$\mathbf{e}_s^2 = \sum_{i=1}^{3} \sum_{j=1}^{3} \mathbf{e}_{ij}^2. \quad (7)$$

From Eqs. (6) and (7) it can be easily seen that: $\mathbf{e}_s = \sqrt{2(u^2 + v^2)}$.

We calculated the components of the monoclinic strains and spontaneous strains tensor as well as the scalar spontaneous strain for the five orthotungstates here studied taking $a_M$, $b_M$, $c_M$, and $\boldsymbol{b}_M$ as a function of pressure from Refs. [16, 17, 21]. The values of $a_T$ and $c_T$ were extrapolated into the pressure regime of the fergusonite phase from its pressure dependence at pressures lower than the transition pressure [16, 17, 21]. For the five studied compounds we had enough experimental data points within the pressure stability range of the scheelite structure for making a good extrapolation. The results obtained for the monoclinic and spontaneous strains of $CaWO_4$ are shown in Fig. 2. It can be seen in Fig. 2b that both longitudinal and shear spontaneous strain components are involved in the symmetry decrease from $I4_1/a$ to $I2/a$. On top of that, the fact that the absolute value of $v$ is slightly higher that the absolute value of $u$, in all the studied pressure range, is consistent with the increase of the $\boldsymbol{b}$ angle upon compression. On the other hand, the spontaneous strain of 3.8% calculated for $CaWO_4$ at the highest pressure of the experiments (8 GPa beyond $P_T$) appears reasonable when compared with those obtained for YNbO$_4$ and LaNbO$_4$ [29, 37] at temperatures of around 100 K under their transition temperature. This fact is in good agreement with Hazen and Finger's conclusions about the inverse relationship of pressure and temperature in ABO$_4$ compounds [38]. On the other hand, in Fig. 2a it can be seen that the monoclinic strains $e_{11}$ and $e_{33}$ are in absolute value considerable larger than $e_{22}$. This fact is coherent with the idea that a displacement of the atomic layers of the scheelite phase along the [100] or [010] directions is involved in the scheelite-to-fergusonite transition [14]. The differences between the diagonal components of the monoclinic strain tensor are a natural consequence of the fact that compression of the fergusonite phase of the compounds typified by CaWO$_4$ is very anisotropic [16, 17, 21].

**VI. Landau theory**

The deviation of the fergusonite structure from the I4$_1$/$a$ symmetry can be expressed by the magnitude of the order parameter η. According to the Landau theory [32] for a second-order transition η is small close to the critical value of the relevant thermodynamic variable (in our case close to $P_T$). Under this assumption, the Gibbs free energy ($G$) of the fergusonite phase relative to the scheelite phase can be expanded in terms of η as $G = \sum_i a_i \mathbf{h}^i$, where $i$ is even since the energy is symmetric with respect to the reversal of polarization direction. The phase transition is driven by the dependence of the lowest-order term on pressure. Then, if we truncate the expansion of $G$ in the second term, we have: $G = a_2(P - P_T)\mathbf{h}^2 + a_4\mathbf{h}^4$. From this equation, the relation between pressure and η can be found by minimizing $G$; i.e. when the condition $\partial G / \partial \mathbf{h} = 0$ is satisfied. It is straightforward to see that this condition is fulfilled if the order parameter has the form: $\mathbf{h} \propto \sqrt{(P - P_T)/P_T} = \mathbf{h}'$, being η' the phenomenological order parameter in Landau's theory. In a ferroelastic transition the spontaneous strain $e_s$ is expected to be closely proportional to the lowest order term to which it couples, namely $\eta^2$. Therefore, if this model of the phase transition behaviour of scheelite-structured orthotungstates is accurate, $P$ should be a quadratic function of $e_s$. In Fig. 2b, it can be easily seen that for $CaWO_4$ this condition is fulfilled. In the other orthotungstates we analyzed, $e_s$ follows the same trend upon compression. In Fig. 3 we plotted $e_s$ versus $\eta'$ for the five compounds here studied. There it can be seen that, within the uncertainty of the results, it exists a linear relationship between the Landau order parameter and the spontaneous strain. In particular, we found that $e_s = 0.0445(8) \eta'$, with a correlation coefficient $r^2 = 0.99$. This fact strongly suggests that the scheelite-to-fergusonite transition here studied is a second-order phase transition. An analogous phase transition was found at low temperature in the isostructural scheelite $CaMoO_4$ [39] giving additional support to our conclusion.

**VII. Summary**

We found that that the magnitude of the spontaneous strain associated with the scheelite-to-fergusonite phase transition of $CaWO_4$, $SrWO_4$, $BaWO_4$, $PbWO_4$, and $EuWO_4$ is proportional to the phenomenological order parameter defined in the Landau theory. This evidence strongly points out that the studied pressure-induced phase transition is a second-order ferroelastic phase transformation. A similar approach can be applied to analyze the character of the pressure-induced phase transitions in other scheelite-structured compounds like the orthomolybdates [40].

**Acknowledgements:** This study was supported by the MCYT of Spain under grant No. MAT2004-05867-CO3-01 and by the Generalitat Valenciana under grant No. ACOMP06/81. The author thanks the financial support of the MCYT and the Universitat de València through the "Ramón y Cajal" program.

**Figure 1:** A perspective view of the scheelite and the fergusonite structures. The large black circles represent the A atoms, the grey circles represent the W atoms, and the small black circles represent the O atoms. The A-O and W-O bonds are shown. In order to better illustrate the differences between the fergusonite and scheelite structures, they were drawn using the structural parameters of CaWO$_4$ at 1.4 GPa (scheelite) and 18.3 GPa (fergusonite).

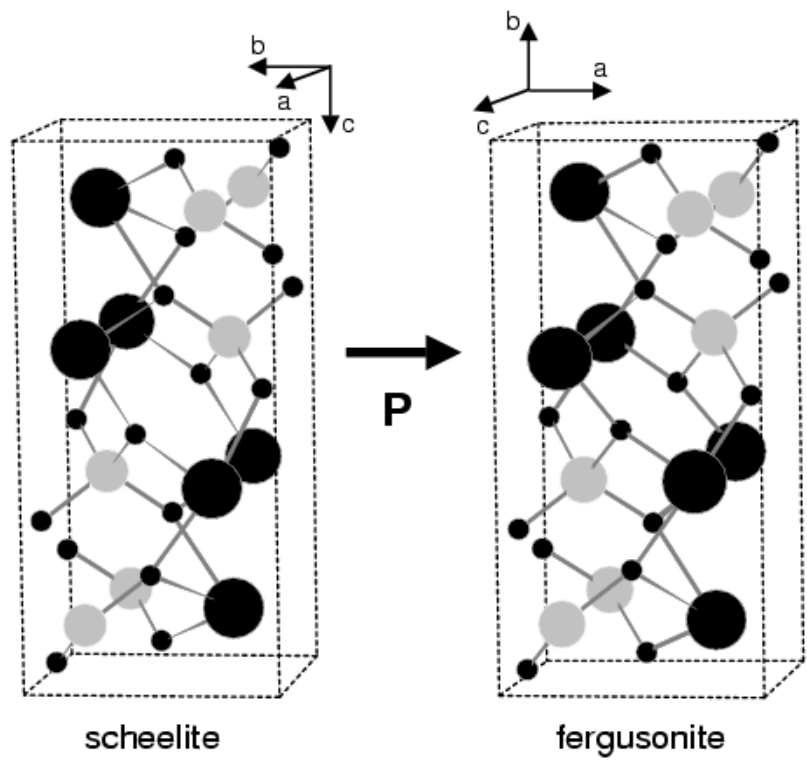

**Figure 2:** Pressure evolution of: **(a)** the monoclinic strains and **(b)** the spontaneous strains of $CaWO_4$.

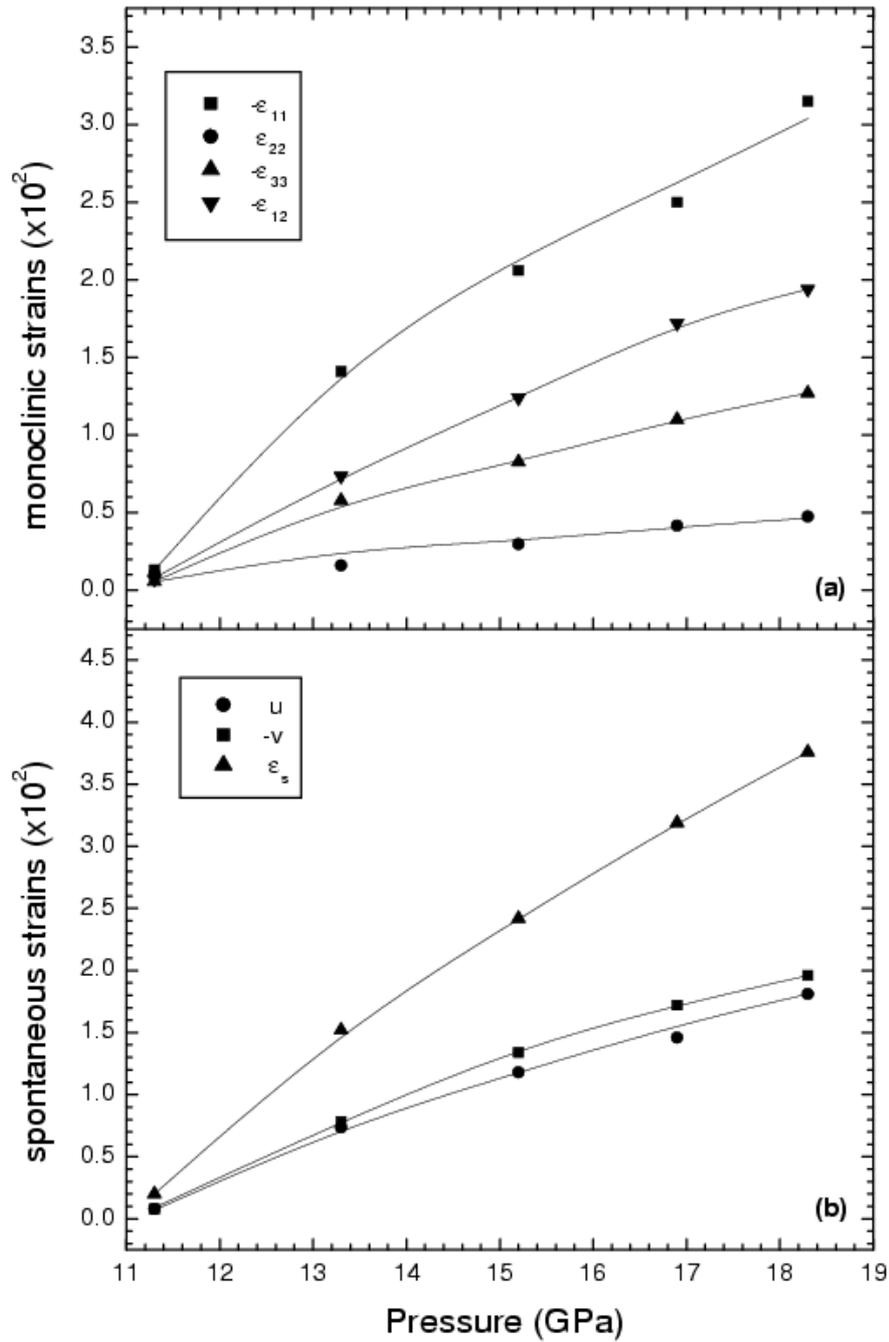

**Figure 3:** Correlation between the spontaneous strain $e_s$ and the Landau order parameter $\eta'$.

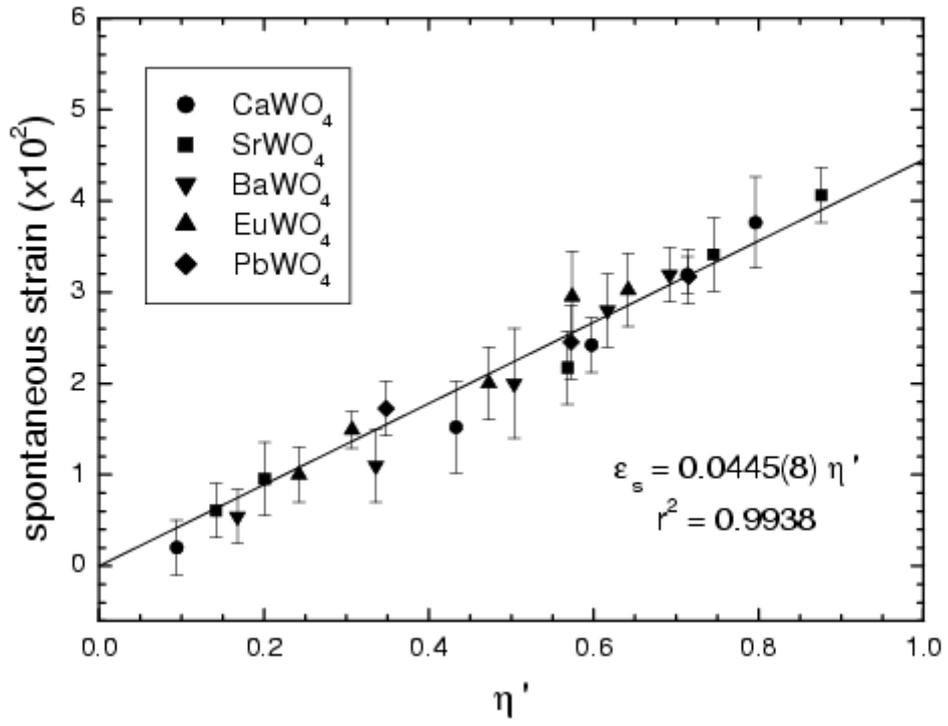